\documentstyle[prc,epsfig,preprint,aps]{revtex}

\begin{document}

\title{Scaling law for the electromagnetic form factors of the proton}

\author{Ding H. Lu$^1$, Shin Nan Yang$^1$ and Anthony W. Thomas$^2$}
\address{$^1$Department of Physics,
    National Taiwan University, Taipei 10617, Taiwan}
\address{$^2$Department of Physics and Mathematical Physics and \break
Special Research Centre for the Subatomic Structure of
    Matter,\break
     University of Adelaide, Australia 5005}
\maketitle

\begin{abstract}
The violation of the scaling law for the electric and magnetic
form factors of the proton is examined within the 
cloudy bag model. We find that the suppression
of the ratio of the electric and magnetic form factors is natural  
in the bag model. The pion cloud  plays a moderate role 
in understanding the recent data from TJNAF.
\end{abstract}

\vspace{1cm}
The description of the electromagnetic structure of the nucleon
requires two independent form factors. The usual Sachs form factors
fully characterize the charge and current distributions inside the nucleon.
Electromagnetic probes interact not only with valence quarks  
confined inside a quark core by nonlinear, gluon dynamics 
but also with the pion field required by chiral symmetry.
A full understanding of the electromagnetic structure of the nucleon 
is of fundamental importance.
  
Historically,  experimental determination of the electric ($G_{Ep}$) 
and magnetic ($G_{Mp}$) 
form factors of the proton was mainly based on the Rosenbluth 
separation~\cite{Rosenbluth50}
of the unpolarized differential cross section data.
The results of various analyses from the early experiments are summarized 
in a simple scaling law, 
\begin{equation}
G_{Ep}(Q^2)=G_{Mp}(Q^2)/\mu_p=G_D(Q^2),
\end{equation}
for momentum transfers, $Q$, up to several GeV.
Here $\mu_p$ is 
proton's  magnetic moment and $G_D(Q^2)$ refers to the standard dipole form.
However, from the Rosenbluth formula one sees 
that at large momentum transfer, 
the electric contribution to the cross section is kinematically
suppressed relative to the magnetic contribution. 
Thus $G_{Ep}$ can not be determined as accurately as $G_{Mp}$ from such
an analysis, especially at large $Q^2$.  
In the literature, the ratios, $\mu_p G_{Ep}(Q^2)/G_{Mp}(Q^2)$, 
obtained from different experiments are
not consistent with each other, within the quoted errors.  

With the advance of polarization technologies and the operation of 
high-duty electron machines,
it is now possible  to drastically reduce the systematic uncertainties  
in this ratio by direct measurement. That is, one can simultaneously 
measure the two components  of the recoil proton polarization, 
$P_T$ and  $P_L$,
using a longitudinally polarized electron beam\cite{Jlab}.
As the transverse component behaves as $P_T\sim G_{Ep}G_{Mp}$,
and the longitudinal component as  $P_{L}\sim G_{Mp}^2$, 
the ratio, $\mu_p G_{Ep}/G_{Mp}$, can be determined 
directly from $P_T/P_L$.

Precise data for the nucleon electromagnetic form factors 
sets a strong constraint
for  various quark models of the nucleon. 
It helps us to develop an understanding of the composite nature of 
the nucleon as well as its long range chiral structure.
In our previous work, we studied the nucleon electromagnetic form factors
in an improved cloudy bag model (CBM)~\cite{CBM98},
where the center-of-mass motion correction
and relativistic effects were treated explicitly. 
As a result, the region of 
validity of the calculation of the 
electric and magnetic form factors in the model was
extended to larger momentum transfer than had previously been possible. 
Here we focus on the ratio, $\mu_p G_{Ep}(Q^2)/G_{Mp}(Q^2)$,
and examine the mechanism for the violation of the scaling law,
Eq. (1).

Let us start with the MIT bag model~\cite{MIT}.
Under the static cavity approximation, the bag surface is spherical and
all valence quarks are in the lowest eigenmode. 
The electric and magnetic form factors for the proton
can be written as
\begin{eqnarray}
G_{Ep}(Q^2)  &=& \int_0^R 4\pi r^2 dr  j_0(Qr)  [g^2(r) + f^2(r)], \\
G_{Mp}(Q^2) &=& 2m_N \int_0^R 4\pi r^2 dr {j_1(Qr)\over Q} [2g(r) f(r)],
\end{eqnarray}
where $R$ is the bag radius, $Q^2=-q^2=\vec{q}^{\,2}$ with $\vec{q}$ 
the three momentum of the photon in the Breit frame,
and $j_l(x)$ refers to the spherical Bessel function. 
For a massless quark, the quark wave functions are given as 
$g(r) = N_Sj_0(\omega_S r/R)$ and
$f(r) = N_S j_1(\omega_S r/R)$,  with $\omega_S=2.04$ 
and $N_S^2 = \omega_S/8\pi R^3j_0^2(\omega_S)(\omega_S-1)$.

At small $Q^2$, the ratio $\mu_p G_{Ep}(Q^2)/G_{Mp}(Q^2)$ is 
determined by the electromagnetic root-mean-squared (r.m.s.) radius,
which is defined as followes: 
$G_{E,M}(Q^2) \sim 1 - Q^2\langle r^2\rangle_{E,M}/6$,
as $Q\rightarrow 0$.
A direct evaluation gives
\begin{eqnarray}
\langle r^2\rangle_{Ep} &=& 4\pi N_S^2\left(R\over \omega_S\right)^5 
{\omega_S^2(2\omega_S^3-2\omega_S^2+4\omega_S-3)\over 
3(2\omega_S^2-2\omega_S+1)}, \\
\langle r^2\rangle_{Mp} &=& {8\pi N_S^2\over 5\mu_p}
\left(R\over\omega_S\right)^6 
{\omega_S^2(8\omega_S^3+10\omega_S^2-20\omega_S+15)\over 
 24(2\omega_S^2-2\omega_S+1)}.
\end{eqnarray}
Note that $\mu_p= R(4\omega_S-3)/12\omega_S(\omega_S -1)$, 
thus the ratio
$\langle r^2\rangle_{Ep}/\langle r^2\rangle_{Mp}$ is independent of $R$.
It results in 1.36 using $\omega_S=2.04$.
This means that the charge r.m.s. radius is considerably larger than 
the magnetic r.m.s. radius.
In other words, 
%
%
$G_{Ep}(Q^2)$ decreases faster than $G_{Mp}(Q^2)$
at small momentum transfers, so the suppression of 
$\mu_p G_{Ep}(Q^2)/G_{Mp}(Q^2)$, as $Q^2$ increases, is a 
natural consequence of the MIT bag model.

The results of our numerical calculations are shown in Fig.~1,
where $R = 0.8$ fm is used. The long-dashed line is the ratio in the 
static MIT bag model. Two lowest curves are respectively electric and 
magnetic form factors of the proton within the MIT bag model.
The experimental data are from the recent measurement at TJNAF~\cite{Jlab}.
Note that the naive MIT bag model for the nucleon is only sensible 
in the region of very small momentum transfers. 
The ratio in a static bag calculation deviates from the data 
very quickly as $Q^2$ increases.
The singularity in the long-dashed curve comes from a
node in $G_{Mp}(Q^2)$, which is related to the sharp bag surface,
and will be extremely sensitive to small admixtures in the nucleon 
ground state wave function. Such admixtures may be induced by residual 
one-gluon exchange or meson exchange interactions\cite{Isgur}.
More sophiscated models, such as the color dielectric and 
soliton bag models~\cite{CDM}, would also help to remove the zero (or at
least move it to larger $Q^2$), 
thus removing the singularity in the ratio.

As the bag model is inherently an independent particle model, the spurious 
center-of-mass motion must be removed in a realistic calculation. 
We have incorporated this correction by constructing 
momentum eigenstates using the Peierls-Thouless projection method\cite{PT}.
Such wave functions are consistent with Galilean translational invariance.
In addition, since the static solution of the quark wave function is spherical, 
the calculated  form factors are trustworthy only at very 
small momentum transfer. To account for relativistic effects
we use the  prescription proposed by Licht and Pagnementa~\cite{LP}. 
The technical details for implementing such corrections can be found
in Ref.~\cite{CBM98}.
The resulting ratio for the quark core of the bag model,
after including the center-of-mass motion corrections and relativistic 
effects, is plotted as the solid curve in Fig.~1. 
The prominent zeros in the form factors are forced out much further 
and do not occur in the momentum region we are studying. 
Compared with the recent data, the improvement over the naive MIT bag model
is significant, in particular, away from the region of
small momentum transfers.  

In the cloudy bag model the physical nucleon is a superposition of a quark core 
and a quark core plus a meson cloud~\cite{CBM84}.
We keep only the dominant pion terms and treat them in 
the one loop approximation.
Inside the pion loop, the intermediate baryon  is
either a $N(939)$ or a $\Delta(1232)$.
Thus the electromagnetic form factors receive contributions from 
the pion current in addition to the direct photon coupling to the 
confined quarks in the core. 
For more details and the explicit expressions for the 
form factors, we refer the readers to Eqs.(36-40) 
and Eqs.(A5,A6) in Ref.~\cite{CBM98}.

In our calculations 
the $\pi NN$ coupling constant is taken to be $f^2_{\pi NN}= 0.0791$.
Other coupling constants are fixed by SU(6) symmetry. 
The resulting ratios in the full calculations are presented in Fig.~2.
Here the solid and long-dashed curves are respectively for massless 
and 10 MeV quarks in the nucleon bag, using a bag radius of $R=0.8$ fm.
Clearly the ratio is insensitive to the quark mass variations.
The addition of the pion cloud for $R=0.8$ fm increases the ratio 
by roughly 20\% towards the recent data.
In the same figure, the dot-dashed and dashed curves have similar meanings 
but with
$R=0.7$ fm. This is about the smallest bag radius which allows
a perturbative treatment of the pion cloud. 
The recent data indicate that a smaller bag radius is prefered. 

One feature of our calculation is that the decline of the ratio 
of the Sachs form factors slows down as $Q^2 > 2$ Gev$^2$.
The corresponding ratio of the Pauli and Dirac form factors 
$Q^2F_{2p}/F_{1p}$ shows a rapid rise at small $Q^2$ (below 2 GeV$^2$)
and seemingly become flat at large $Q^2$.
This is consistent with the prediction of pQCD, which claims
$F_{1p}\sim 1/Q^4$ and $F_{2p}\sim 1/Q^6$, 
but is not obvious in other nucleon models~\cite{CQM,DQ,VMD,Skyrme}. 

In summary, we have studied the ratio, $\mu_p G_{Ep}(Q^2)/G_{Mp}(Q^2)$,
in an improved quark model.
We find that the sharp decline of the ratio as $Q^2$ increases 
is natural in a quark bag model, so the empirical scaling law for the 
nucleon electromagnetic form factors is unavoidably violated. 
The deviations of the ratio, $\mu_p G_{Ep}(Q^2)/G_{Mp}(Q^2)$ 
from unity can be understood semi-quantitatively 
within the cloudy bag model.

This work was supported in part by the National Science Council
of ROC under grant No. NSC-89-2112-M002-038 and 
the Australian Research Council.

\begin{figure}[t]
\vspace{1.5cm}
\centering{\
\epsfig{file=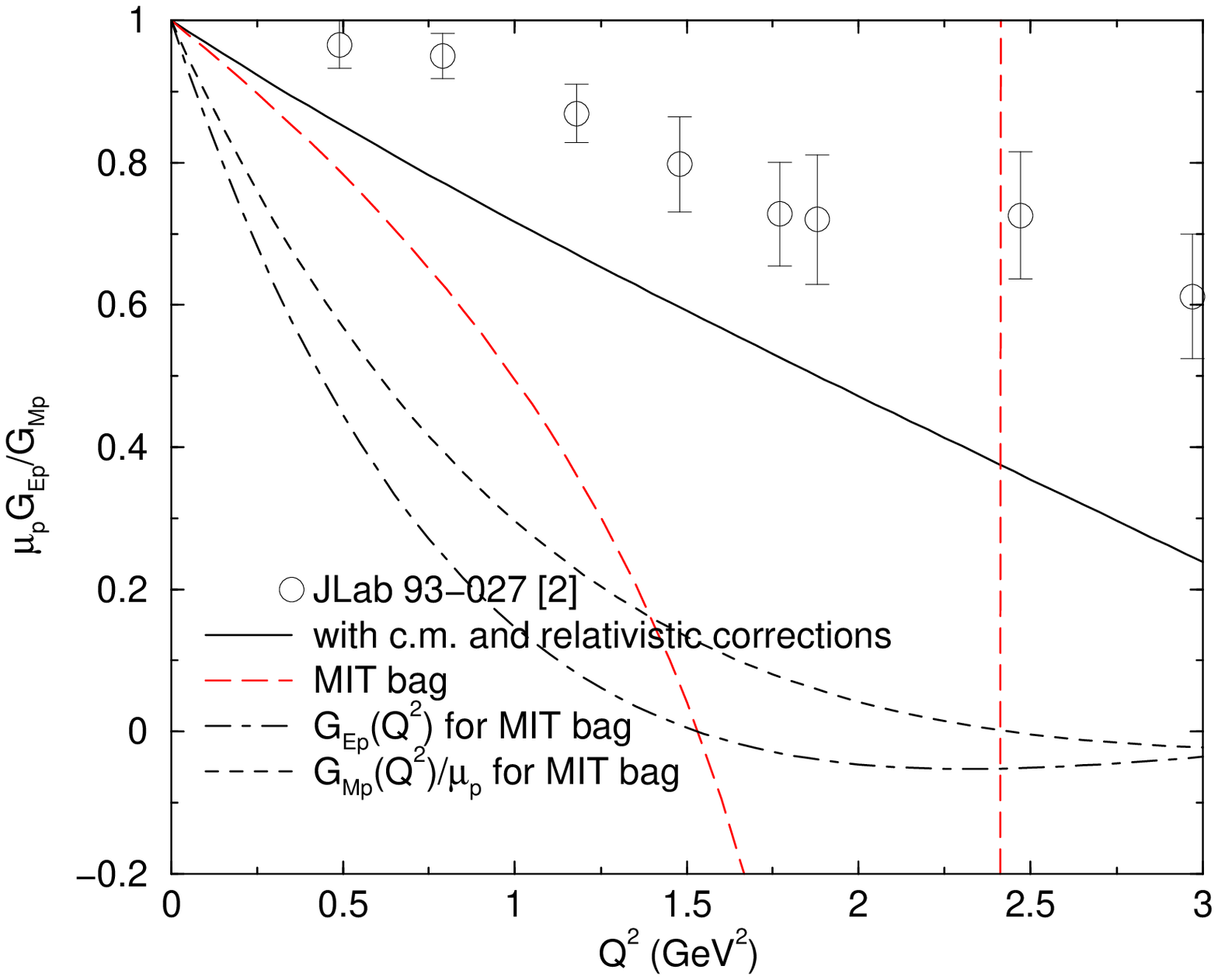,height=8.0cm,width=10cm}
\vspace*{1.0cm}
\caption{The ratio, $\mu_p G_{Ep}(Q^2)/G_{Mp}(Q^2)$,  
calculated from the quark core of the bag model.  
The solid curve is the result after the center-of-mass motion correction 
and a semi-classical relativistic correction. 
The data are from the recent measurement at TJNAF~[2].}
\label{fig1.eps}}
\end{figure}

\begin{figure}[t]
\vspace{1.5cm}
\centering{\
\epsfig{file=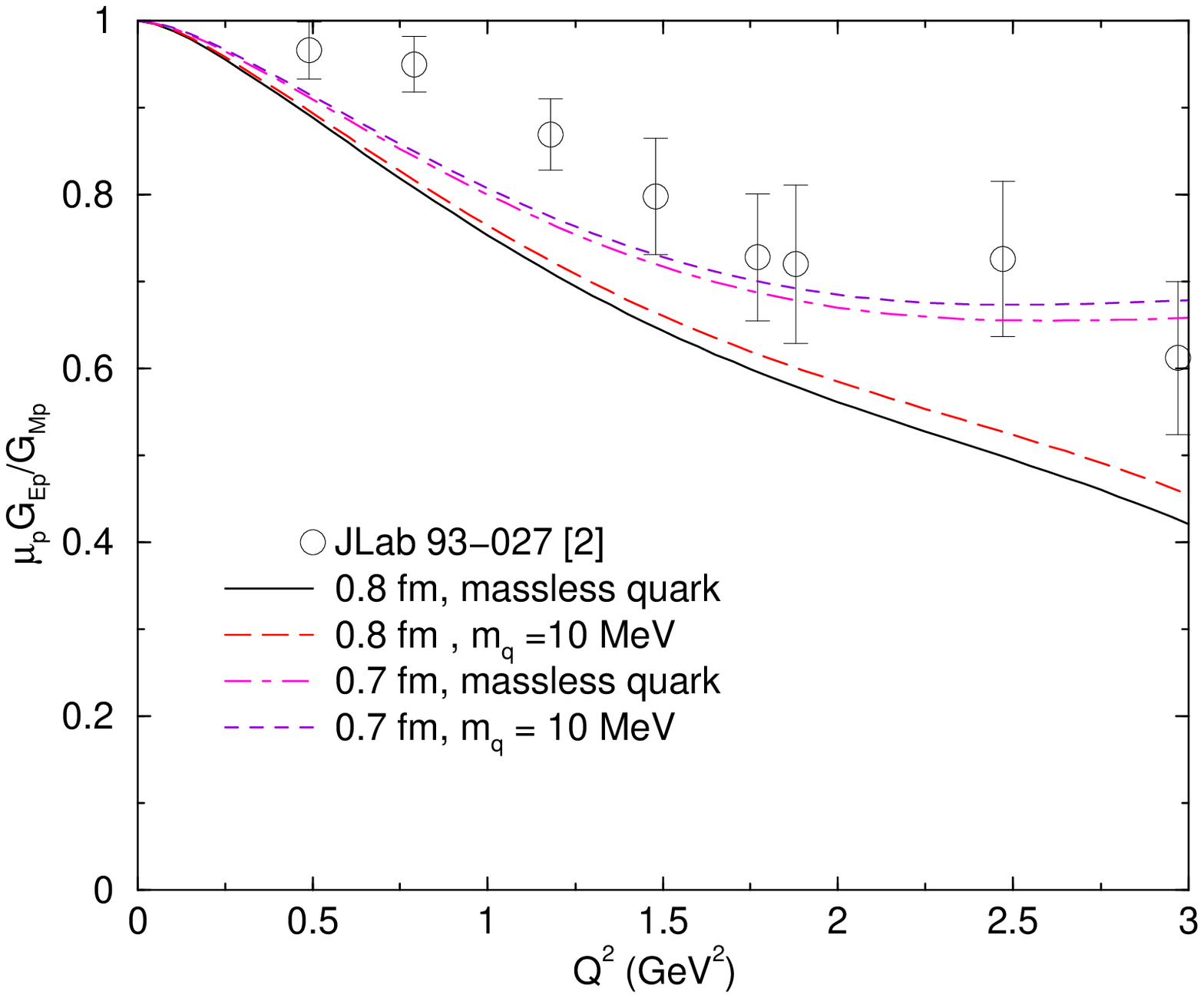,height=8cm,width=10cm}
\vspace*{1.0cm}
\caption{The ratio, $\mu_p G_{Ep}(Q^2)/G_{Mp}(Q^2)$, from full calculations 
in the improved cloudy bag model.
Results with different bag radius and quark masses are also presented.
The data are from the recent measurement at TJNAF~[2].}
\label{fig2.eps}}
\end{figure}


\begin{references}
\bibitem{Rosenbluth50} M. N. Rosenbluth, Phys. Rev. {\bf 79}, 615 (1950).
\bibitem{Jlab} M. K. Jones {\em et al.}, submitted to Phys. Rev. Lett. 
    (Oct. 1999) and nucl-ex/9910005.
\bibitem{CBM98} D. H. Lu, A. W. Thomas, and A. G. Williams, 
    Phys. Rev. C {\bf 57}, 2628 (1998); {\em ibid.} {\bf 55}, 3108 (1997).
\bibitem{MIT} A. Chodos, {\em et al.}, Phys. Rev. D {\bf 9}, 3471 (1974);
    A. Chodos {\em et al., ibid.} {\bf 10}, 2599 (1974); 
    T. A. DeGrand {\em et al., ibid.} {\bf  12}, 2060 (1975).
\bibitem{PT} R. E. Peierls and D. J. Thouless, 
    Nucl. Phys. {\bf 38}, 154 (1962).
\bibitem{LP}A. L. Licht and A. Pagnmenta, Phys. Rev. D {\bf 2}, 1156 (1970).
\bibitem{CBM84} A. W. Thomas, Adv. Nucl. Phys. {\bf 13}, 1 (1984);
	G. A. Miller, Int. Rev. Nucl. Phys. {\bf 2}, 190 (1984).
\bibitem{Isgur} S. Capstick and N. Isgur, Phys. Rev. D {\bf 34}, 2809 (1986);
	P. Geiger and N. Isgur, {\em ibid.} {\bf 55}, 299 (1997).
\bibitem{CDM} M. C. Birse, Prog. Part. Nucl. Phys. {\bf 25}, 1 (1990);
 L. Wilets, {\em Non-Topological Solitons} (World Scientific, Singapore, 1989).
\bibitem{CQM} P. L. Chung and F. Coester, Phys. Rev. D {\bf 44}, 229 (1991);
	S. Capstick and B. Keister. {\em ibid.} {\bf 51}, 3598 (1995);
	F. Cardarelli, E. Pace, G. Salme, and S. Simula, 
	Nucl. Phys. A {\bf 623}, 361 (1997).
\bibitem{DQ} 
	P. Kroll, M. Schurmann, and W. Schweiger,
	Z. Phys. A {\bf 338}, 339 (1991).
\bibitem{VMD} P. Mergell, U.G. Meissner, and D. Drechsel,
		Nucl. Phys. A {\bf 596}, 367 (1996);
	M. F. Gari and W. Krumpelmann, Z. Phys. A {\bf 322}, 689 (1985);
	G. Hoehler {\em et al.}, Nucl. Phys. B {\bf 114}, 505 (1976).
\bibitem{Skyrme}G. Holzwarth, Z. Phys. A {\bf 356}, 339 (1996); 
	X. Ji, Phys. Lett. B {\bf 254}, 456 (1991).
\end{references}
\end{document}